\documentclass[english,superscriptaddress,nofootinbib,preprintnumbers]{revtex4}
\usepackage[T1]{fontenc}
\usepackage[latin1]{inputenc}
\usepackage{amsmath}
\usepackage{graphicx}
\usepackage{amssymb}

\providecommand{\tabularnewline}{\\}

\makeatletter\makeatletter\makeatother\makeatother\makeatother

\usepackage{babel}

\usepackage{ifpdf}

\usepackage{epstopdf}

\graphicspath{{./}}

\usepackage{babel}

\usepackage{babel}

\begin{document}

\preprint{HD-THEP-09-07}

\title{Cosmic Parallax as a probe of late time anisotropic expansion}

\author{Claudia Quercellini}

\email{claudia.quercellini@uniroma2.it}

\affiliation{Università di Roma Tor Vergata, Via della Ricerca Scientifica 1,
00133 Roma, Italy}

\author{Paolo Cabella}

\email{cabella@roma2.infn.it}

\affiliation{Università di Roma Tor Vergata, Via della Ricerca Scientifica 1,
00133 Roma, Italy}

\author{Luca Amendola}

\email{amendola@mporzio.astro.it}

\affiliation{INAF/Osservatorio Astronomico di Roma, V. Frascati 33, 00040 Monteporzio
Catone, Roma, Italy}

\author{Miguel Quartin}

\email{quartin@mporzio.astro.it}

\affiliation{INAF/Osservatorio Astronomico di Roma, V. Frascati 33, 00040 Monteporzio
Catone, Roma, Italy}

\affiliation{Institut für Theoretische Physik, Universität Heidelberg, Philosophenweg
16, 69120 Heidelberg, Germany}

\author{Amedeo Balbi}

\email{balbi@roma2.infn.it}

\affiliation{Università di Roma Tor Vergata, Via della Ricerca Scientifica 1,
00133 Roma, Italy}

\date{\today{}}
\begin{abstract}
Cosmic parallax is the change of angular separation between pair of
sources at cosmological distances induced by an anisotropic expansion.
An accurate astrometric experiment like Gaia could observe or put
constraints on cosmic parallax. Examples of anisotropic cosmological
models are Lemaitre-Tolman-Bondi void models for off-center observers
(introduced to explain the observed acceleration without the need
for dark energy) and Bianchi metrics. If dark energy has an anisotropic
equation of state, as suggested recently, then a substantial anisotropy
could arise at $z\lesssim1$ and escape the stringent constraints
from the cosmic microwave background. In this paper we show that such
models could be constrained by the Gaia satellite or by an upgraded
future mission. 
\end{abstract}
\maketitle

\section{Introduction}

\label{intro}

In recent times, there has been a resurgent interest towards anisotropic
cosmologies, classified in terms of Bianchi solutions to general relativity.
This has been mainly motivated by hints of anomalies in the cosmic
microwave background (CMB) distribution observed on the full sky by
the WMAP satellite \cite{2004PhRvD..69f3516D,2004ApJ...609...22V,2005MNRAS.356...29C,2004ApJ...605...14E}.
While the CMB is very well described as a highly isotropic (in a statistical
sense) Gaussian random field, recent analyses have shown that local
deviations from Gaussianity in some directions (the so called cold
spots, see \cite{2005MNRAS.356...29C}) cannot be excluded at high
confidence levels. Furthermore, the CMB angular power spectrum extracted
from the WMAP maps has a quadrupole power which appears significantly
lower than expected from the best-fit cosmological model \cite{2004MNRAS.348..885E}.
Several explanations for this anomaly have been proposed (see e.g.
\cite{2003PhLB..574..141T,2003JCAP...09..010C,2003PhRvD..67j3509D,2007PhRvD..76f3007C,2007PhRvD..76h3010G})
including the fact that the universe is expanding with different velocities
along different directions. While deviations from homogeneity and
isotropy are constrained to be very small from cosmological observations,
these usually assume the non-existence of anisotropic sources in the
late universe. Conversely, as suggested in ~\cite{2008JCAP...06..018K,2008ApJ...679....1K,2006PhRvD..74d1301B,2006PhRvD..73f3502C,2008arXiv0812.0376C},
dark energy with anisotropic pressure acts as a late-time source of
anisotropy. Even if one considers no anisotropic pressure fields,
small departures from isotropy cannot be excluded, and it is interesting
to devise possible strategies to detect them.


The effect of assuming an anisotropic cosmological model on the CMB
pattern has been studied by \cite{1973MNRAS.162..307C,1985MNRAS.213..917B,1995A&A...300..346M,1996A&A...309L...7M,1996PhRvL..77.2883B,1997PhRvD..55.1901K}.
The Bianchi solutions describing the anisotropic line element were
treated as small perturbations to a Friedmann-Robertson-Walker (FRW)
background. Such early studies did not consider the possible presence
of a non-null cosmological constant or dark energy and were upgraded
recently by \cite{2006MNRAS.369.1858M,2006ApJ...644..701J}.

One difficulty of the anisotropic models that have been shown to fit
the large-scale CMB pattern is that they have to be produced according
to very unrealistic choices of the cosmological parameters. For example,
the Bianchi VIIh template used in \cite{2006ApJ...644..701J} requires
an open universe, an hypothesis which is excluded by most cosmological
observations. An additional problem is that an inflationary phase
-- required to explain a number of feature of the cosmological model
-- isotropizes the universe very efficiently, leaving a residual anisotropy
that is negligible for any practical application. These difficulties
vanish if an anisotropic expansion takes place only well after the
decoupling between matter and radiation, for example at the time of
dark energy domination ~\cite{2008JCAP...06..018K,2008ApJ...679....1K,2006PhRvD..74d1301B,2006PhRvD..73f3502C,2008arXiv0812.0376C}.

The effect of cosmic parallax~\cite{2008arXiv0809.3675Q} has been
recently proposed as a tool to assess the presence of an anisotropic
expansion of the universe. It is essentially the change in angular
separation in the sky between far-off sources, due to an anisotropic
expansion. This all-sky change in separations can be used as a tracer
of anisotropic behaviour of the spacetime metric. This effect has
been investigated in the context of Lemaître-Tolman-Bondi (LTB) models
with off-center observers~\cite{2008arXiv0809.3675Q}. In this paper
we study the cosmic parallax in Bianchi I metrics (see also \cite{2009arXiv0905.3727F}
for ellipsoidal universes). We will show that, since the cosmic parallax
traces the geodesic of the metric to the present time, it can be used
to constrain the late anisotropic behaviour induced, for example,
by the above mentioned anisotropic dark energy models. This makes
it a valuable tool with respect to primary CMB anisotropies which
are frozen at $z\sim1000$.

While finalizing this paper another work analysing the cosmic parallax
in Bianchi I models appeared \cite{2009arXiv0905.3727F}. We will
discuss the main differences with this work later on.

\section{Cosmic parallax in Bianchi I}

\label{cp} We consider a class of homogeneous and anisotropic models
where the line element is of the Bianchi I type, \begin{equation}
ds^{2}=-dt^{2}+a^{2}(t)dx^{2}+b^{2}(t)dy^{2}+c^{2}(t)dz^{2}.\label{metric}\end{equation}
 The expansion rates in the three Cartesian directions $x$, $y$
and $z$ are defined as $H_{X}=\dot{a}/a$, $H_{Y}=\dot{b}/b$ and
$H_{Z}=\dot{c}/c$, where the dot denotes the derivative with respect
to coordinate time. In these models they differ from each other, but
in the limit of $H_{X}=H_{Y}=H_{Z}$ the flat FRW isotropic expansion
is recovered. Among the Bianchi classification models the type I exhibits
flat geometry and no overall vorticity; conversely, shear components
$\Sigma_{X,Y,Z}=H_{X,Y,Z}/H-1$ are naturally generated, where $H$
is the expansion rate of the average scale factor, related to the
volume expansion as $H=\dot{A}/A$ with $A=(abc)^{1/3}$.


Cosmic parallax is the temporal change of angular separations between
distant sources in the sky caused by large scale anisotropic expansion
\cite{2008arXiv0809.3675Q}. The sources are assumed to trace the
evolution of the cosmic expansion (see for example also \cite{2009arXiv0903.3402D});
since the parallax induced by peculiar velocity is randomly distributed,
it can be averaged out of a large sample and, in addition, decreases
with distance from the observer~\cite{2008arXiv0809.3675Q}.

For an off-centre observer in a LTB model the cosmic parallax is a
pure dipole signal in the sky that might be affected by systematic
noises like the observer's peculiar velocity and acceleration (the
latter induces aberration changes), even though an observational strategy
using different source samples at different redshifts (say, within
and outside the void) would help to discriminate between them. In
homogeneous and anisotropic models like Bianchi I we expect the signal
to have a different angular distribution, hence being even more predictive.

\begin{figure}[t]
 \includegraphics[width=7cm]{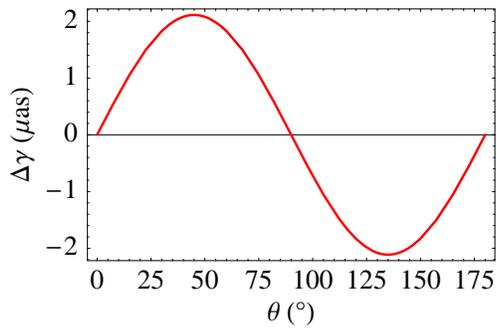}

\caption{Cosmic parallax in Bianchi I models as a function of $\theta$ for
$\phi=\Delta\phi=0$ and $\Delta\theta=90^{o}$ (this setting corresponds
to the (X,Z) plane), $h_{x}=0.71$, $h_{y}=0.725$, $h_{z}=0.72$,
i.e. $\Sigma_{0X}=-0.012$ and $\Sigma_{0Y}=0.009$). The time interval
is $\Delta T=10$yrs.}

\label{bianthe} 
\end{figure}

\begin{figure}[t]
 \includegraphics[width=7cm]{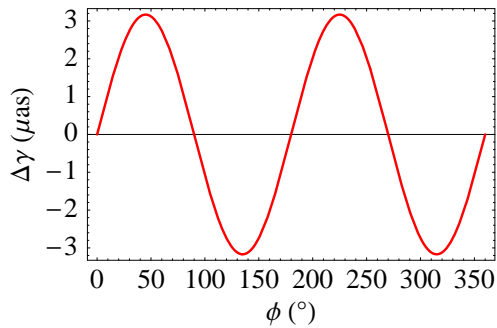}

\caption{Cosmic parallax in Bianchi I models as a function of $\phi$ for $\theta=90^{o}$,
$\Delta\theta=0$ and $\Delta\phi=90^{o}$ (this setting corresponds
to the (X,Y) plane), $h_{x}=0.71$, $h_{y}=0.725$, $h_{z}=0.72$,
i.e. $\Sigma_{0X}=-0.012$ and $\Sigma_{0Y}=0.009$). The time interval
is $\Delta T=10$yrs. }

\label{bianphi} 
\end{figure}

Let us consider two sources $A$ and $B$ in the sky located at physical
distance from us observers $\vec{O}_{[A,B]}=(X,Y,Z)_{[A,B]}=(R\sin{\theta}\cos{\phi},R\sin{\theta}\sin{\phi},R\cos{\theta})_{[A,B]}\,$,
where $R=\sqrt{X^{2}+Y^{2}+Z^{2}}$ and $(\theta,\phi)$ are spherical
angular coordinates. Their angular separation on the celestial sphere
reads \begin{eqnarray}
\vec{OA}\cdot\vec{OB}=\cos{\gamma}=\cos{\theta_{A}}\cos{\theta_{B}}+\sin{\theta_{A}}\sin{\theta_{B}}\cos{\Delta\phi},\label{gamma}\end{eqnarray}
 with $\Delta\phi=(\phi_{A}-\phi_{B})$. From now on we will mark
spatial separations and temporal variations with $\Delta$ and $\Delta_{t}$
symbols, respectively. If $\Delta_{t}\gamma\ne0$ then a cosmic parallax
arises. In a homogeneous and isotropic model (like FRW) the geodesic
are radial and sources are subjected to the same radial expansion
rate, keeping their angular separation constant. On the other hand,
if the expansion is anisotropic their spherical coordinates change
dissimilarly in time leading to a modification of their angular separation:
\begin{eqnarray}
-\sin{\gamma}\Delta_{t}\gamma & = & \sin{\theta_{A}}\cos{\theta_{B}}(\Delta_{t}\theta_{B}\cos{\Delta\phi}-\Delta_{t}\theta_{A})+\cos{\theta_{A}}\sin{\theta_{B}}(\Delta_{t}\theta_{A}\cos{\Delta\phi}-\Delta_{t}\theta_{B})\label{dgamma}\\
 & + & \sin{\theta_{A}}\sin{\theta_{B}}\sin{\Delta\phi}(\Delta_{t}\phi_{B}-\Delta_{t}\phi_{A}).\nonumber \end{eqnarray}
 In the limit $\Delta_{t}\phi_{A}=\Delta_{t}\phi_{B}=\phi_{A}=\phi_{B}=0$
the relative motion is constrained on the (X,Z) plane and the cosmic
parallax reduces to $(\Delta_{t}\theta_{A}-\Delta_{t}\theta_{B})$
(see Fig.~\ref{bianthe}). Similarly, on the (X,Y) plane the signal
is $(\Delta_{t}\phi_{A}-\Delta_{t}\phi_{B})$ (see Fig.~\ref{bianphi}).
Both in Fig.~\ref{bianthe} and \ref{bianphi} we allowed the shear
parameters at present to appreciably deviate from 0. This explains
why the cosmic parallax is few orders of magnitude larger than the
one in \cite{2009arXiv0905.3727F}. The main motivation for this will
be presented in the first paragraph of Section~\ref{DE}.

The signal is in general a combination of both the anisotropic expansion
of the sources themselves and the change in curvature induced by the
shear on the photon path from the emission to the observer. In inhomogeneous
and/or anisotropic models photons follow trajectories that, in general,
are not radial. However, while in LTB models this effect on the cosmic
parallax is enhanced by inhomogeneity (although a FRW description
of null geodesic has been shown to be fairly good approximation \cite{2008arXiv0809.3675Q}),
in Bianchi I models we consider in this paper the geodesic bending
for a single source amounts at most to about 7$\%$ (see Appendix
A), which allow us to adopt the straight geodesics approximation.

The spherical angular coordinates are related to the Cartesian coordinates
via $\phi=\arctan{(Y/X)}$ and $\theta=\arccos{(Z/\sqrt{X^{2}+Y^{2}+Z^{2}})}$.
Therefore their time evolution can be written as \begin{eqnarray}
\Delta_{t}\phi & = & \frac{XY}{X^{2}+Y^{2}}(H_{0Y}-H_{0X})\Delta t\label{dang}\\
\Delta_{t}\theta & = & \frac{Z\, R^{-2}}{\sqrt{X^{2}+Y^{2}}}\Big[X^{2}(H_{0X}-H_{0Z})+Y^{2}(H_{0Y}-H_{0Z})\Big]\Delta t,\label{dangb}\end{eqnarray}
 where we made use of the Hubble law in the three cartesian directions
$\Delta_{t}X,Y,Z=X,Y,Z\cdot H_{0X,Y,Z}\Delta t$, valid for small
time span $\Delta t$ (which we assume of the order of decades) relatively
to the cosmic time.

In spherical coordinates Eqs.~(\ref{dang}-\ref{dangb}) can be expressed
as \begin{eqnarray}
\Delta_{t}\phi & = & \frac{\sin{2\phi}}{2}(\Sigma_{0Y}-\Sigma_{0X})H_{0}\Delta t\label{dang2}\\
\Delta_{t}\theta & = & \frac{\sin{2\theta}}{4}\Big[(3(\Sigma_{0X}+\Sigma_{0Y})+\cos{2\phi}(\Sigma_{0X}-\Sigma_{0Y})\Big]H_{0}\Delta t,\label{dang2b}\end{eqnarray}
 where $\Sigma_{0X,Y,Z}$ are the shear components at present as defined
at the beginning of this section, satisfying the transverse condition
$\Sigma_{0X}+\Sigma_{0Y}+\Sigma_{0Z}=0$.

Equations~(\ref{dang2}-\ref{dang2b}) describe a pure quadrupole
signal in the $\phi$ and $\theta$ coordinate, respectively. This
functional form of the signal is exactly the same as the one expected
for the first non-vanishing multipole expansion of the CMB large scale
relative temperature anisotropies in Bianchi I model~\cite{1995A&A...300..346M}
(remember we are neglecting all peculiar velocities, including our
own). By combining them into the full spherical distance formula~(\ref{dgamma})
the resulting cosmic parallax signal obviously exhibits a more complicated
shape depending on the location on the sky.

Considering two sources with an initial angular separation such that
$(\theta,\phi)_{B}=(\theta,\phi)_{A}+\Delta(\theta,\phi)$ and substituting
Eqs.~(\ref{dang2}-\ref{dang2b}) in Eq.~(\ref{dgamma}) we gain
the full expression for the cosmic parallax in spherical coordinates
$\Delta_{t}\gamma=\Delta_{t}\gamma(\theta,\phi,\Delta\theta,\Delta\phi,\Sigma_{0X},\Sigma_{0Y},H_{0},\Delta t)$,
where the only further conjecture is that $H$ does not vary in $\Delta t$.
At first order, this seems reasonable for the time intervals under
consideration. Notice that the signal turns out to be independent
on the redshift: source pairs along the same line of sight undergo
the same temporal change in their angular separation. This means that
aligned quasars would stay aligned. In general of course the number
density of quasars will change so that the number counts should show
some level of anisotropy. This could provide an additional constraint
on anisotropic expansions: we will discuss briefly this possibility
in Sect. V.

If there were no anisotropies present at last scattering of course
a late time anisotropic expansion would marginally affect the CMB
via a late time direction dependent integrated Sachs- Wolf. Notice
that the positional shift of the sources themselves is a completely
different signal with respect to the bending of light ray during propagation
time.

\begin{figure*}[t]
 \includegraphics[width=14cm]{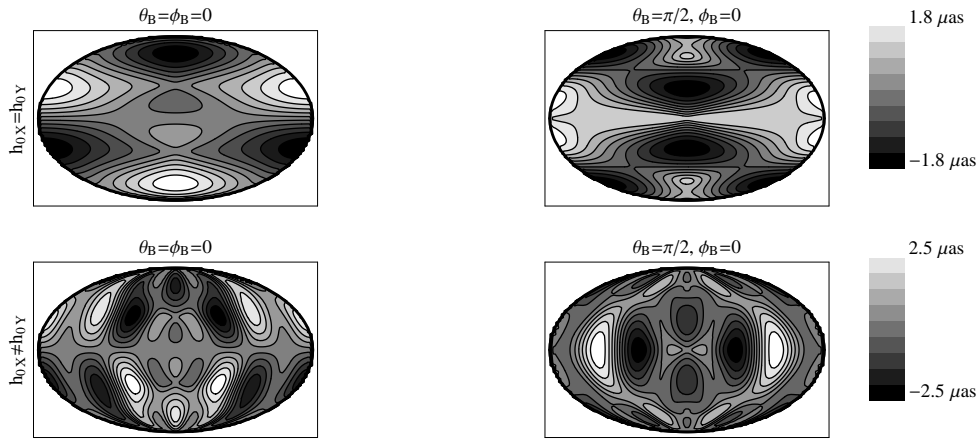}

\caption{Mollweide contour plot for cosmic parallax in Bianchi I models for
one source fixed at two different location in the sky. Upper panels
show the signal for ellipsoidal models ($h_{0Z}=0.72$ and $h_{0X}=h_{0Y}=0.71$),
while in lower panels $h_{0Y}=0.725$. Lighter colours correspond
to higher signal and on the horizontal and vertical axes angular coordinates
vary in the range $\phi:[0,2\pi]$ and $\theta:[0,\pi]$, respectively.
The time interval is $\Delta T=10$yrs.}

\label{moll} 
\end{figure*}

In Fig.~\ref{moll}, Mollweide projections on the sky of the cosmic
parallax signal with respect to a fixed source located on the north
pole and on the (X,Y) plane are shown. As expected, when the source
is at an equatorial position the symmetry with respect to the (X,Y)
plane is preserved, while when the source is at the north pole a symmetry
with respect to the (X,Z) plane emerges. In a FRW universe the components
of the shear simultaneously vanish and so does the cosmic parallax.

\section{Cosmic parallax forecastings}

\begin{table}[t!]
 \centering 
\label{gaia} \begin{tabular}{@{}llll@{}}
\hline 
 &  &  & \tabularnewline
Experiment  & $N_{s}$  & $\sigma_{acc}$  & $\Delta t$ \tabularnewline
 &  &  & \tabularnewline
\hline 
 &  &  & \tabularnewline
Gaia  & 500,000  & 50$\mu$as  & 5yrs \tabularnewline
 &  &  & \tabularnewline
Gaia+  & 1,000,000  & 5$\mu$as  & 10yrs \tabularnewline
 &  &  & \tabularnewline
\hline
\end{tabular}

\caption{Specifications adopted for Gaia-like and Gaia+ experiments, where
$N_{s}$ is the total number of sources, $\sigma_{acc}$ is the experimental
astrometric accuracy and $\Delta t$ is the time interval between
two measurements. }

\end{table}

\label{simul} As a next step, we would like to give an insight on
whether accurate future satellite astrometry mission will be able
to put constraints on the anisotropy parameters that are competitive
with CMB quadrupole constraints \cite{1995A&A...300..346M}. An astrometry
mission like Gaia will detect around 500,000 quasars in its 5 years
flight time with positional error $\sigma_{acc}=$10-200 $\mu$as
\cite{Bailer:2004,Lindegren:2008}. Attributing to a Gaia-like experiment
the capability of detecting the quasar angular positions at two different
time separated by $\Delta t\approx10$yrs (i.e. conceiving the possibility
of two separated missions or just a longer one) we can adopt its instrumental
characteristics to perform a Fisher matrix analysis. In these Bianchi
I models the cosmic parallax signal depends on four parameters: the
average Hubble function at present, the time span and the two Hubble
normalized anisotropy parameters at present. However, for the allowed
range of values, contours in the $(\Sigma_{0X},\Sigma_{0Y})$ frame
do not depend on the value of $H_{0}$. Stretching the time interval
between the two measurements or improving the instrumental accuracy
would instead have an impact on the final constraints. In order to
analyse these dependencies we make use of the Fisher formalism, namely
the Fisher Matrix defined as \begin{equation}
F_{i,j}=\sum_{l}\frac{\partial\Delta_{t}\gamma_{(l)}}{\partial\Sigma_{0i}}\frac{1}{\sigma_{acc}^{2}}\frac{\partial\Delta_{t}\gamma_{(l)}}{\partial\Sigma_{0j}},\label{fish}\end{equation}
 where all separations are taken with respect to a reference source
and index $l$ runs from $2$ up to the number of quasars $N_{s}$
to take into account the spherical distances to all other sources.
In fact, one should notice that the Gaia accuracy positional errors
are obtained having already averaged over 2$N_{s}$ coordinates.

We simulated a catalogue of up to 1,000,000 quasars with angular positions
$(\theta,\phi)$ randomly generated from a uniform distribution on
the celestial sphere. We then used the covariance matrix $C_{ij}=(F_{ij})^{-1}$
to construct the error ellipses with $1$ and $2\sigma$ contours
in the $(\Sigma_{0X},\Sigma_{0Y})$ plane.The positional accuracies
should have a mild dependence on the magnitude. The quasars are expected
to have magnitudes ranging from 12 to 20 and, correspondingly, accuracies
from 10 down to 200 $\mu$as, as pointed out in \cite{Lindegren:2008}.
We could have weighted our non-redshift dependent signal with accuracies
that are function of magnitude. However, for simplicity we adopt a
single representative average accuracy of $\sigma_{acc}=$50 $\mu$as;
it is immediate to rescale the final errors to a different accuracy.
We also perform the calculation for an enhanced Gaia-like mission
dubbed as Gaia+ (see specifications in Table~\ref{gaia}).

\begin{figure}[t]
 \includegraphics[width=8.3cm]{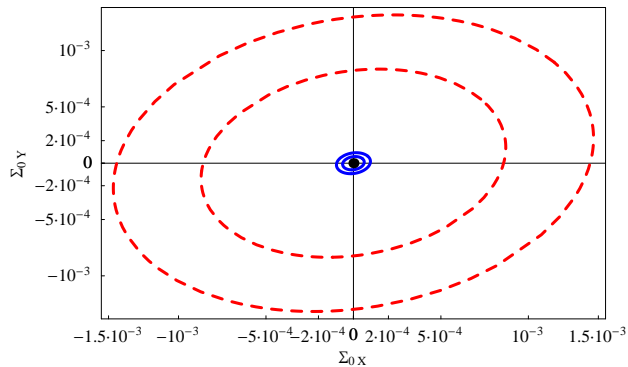}

\caption{Fisher contours of Cosmic parallax for Gaia and Gaia+ specifications
(dashed and solid lines, respectively). The double contours identify
$1\sigma$ and $2\sigma$ regions for $\Delta T=10$yrs.}

\label{cont1} 
\end{figure}

The Fisher error ellipse are shown in Fig~\ref{cont1}; the constraints
turn out to be of the same order of magnitude of the CMB limits on
the shear at decoupling. The 1$\sigma$ errors on $\Sigma_{0X}$ and
$\Sigma_{0Y}$ are $8.3\cdot10^{-4}$ and $6\cdot10^{-5}$ for Gaia
and Gaia+, respectively. Although our null hypothesis was chosen to
be the friedmannian isotropic expansion ($\Sigma_{0X}=\Sigma_{0Y}=0$
and $H_{0}=72$ km/s/Mpc), due to the linear dependence of the signal
on the shear parameters, a change of the fiducial model corresponds
to a simple translation of the same ellipse in the frame. Gaia data
processing is incredibly complex and the experimental covariance matrix
will probably be at the end non-diagonal. However, the Gaia collaboration
have not provided yet a quantification of these correlations and giving
a formal status to it is beyond the scope of this paper.

\section{Cosmic parallax induced by dark energy: an example }

\label{DE} The CMB provides very tight constraints on Bianchi models
at the time of recombination~\cite{1996PhRvL..77.2883B,1997PhRvD..55.1901K,1995A&A...300..346M}
of order of the quadrupole value, i.e. $\sim10^{-5}$. Usually, in
standard cosmologies with a cosmological constant the anisotropy parameters
scale as the inverse of the comoving volume. This implies an isotropization
of the expansion from the recombination up to the present, leading
to the typically derived constraints on the shear today, namely $\sim10^{-9}\div10^{-10}$
(resulting in a cosmic parallax signal of order $10^{-4}\mu$as).
However, this is only true if the aforementioned parameters are monotonically
decreasing functions of time, that is if the anisotropic expansion
is not generated by any anisotropic source arising after decoupling,
e.g. vector fields representing anisotropic dark energy ~\cite{2008ApJ...679....1K}.

Motivated by this, we apply the cosmic parallax to a specific anisotropic
phenomenological dark energy model in the framework of Bianchi I models~\cite{2008ApJ...679....1K,2008JCAP...06..018K}
(we refer to these papers for details). This description allows semi-analytical
calculations and represents in a fairly conservative approach more
complicated anisotropic models. Here the anisotropic expansion is
caused by the anisotropically stressed dark energy fluid whenever
its energy density contributes to the global energy budget. If the
major contributions to the overall budget come from matter and dark
energy, as after recombination, their energy-momentum tensor can be
parametrized as: \begin{eqnarray}
T_{(m)\nu}^{\mu} & = & \mbox{diag}(-1,w_{m},w_{m},w_{m})\rho_{m}\\
T_{({\rm DE)\nu}}^{\mu} & = & \mbox{diag}(-1,w,w+3\delta,w+3\gamma)\rho_{{\rm DE}},\label{eq:Tmunu-de}\end{eqnarray}
 respectively, where $w_{m}$ and $w$ are the equation of state parameters
of matter and dark energy and the skewness parameters $\delta$ and
$\gamma$ can be interpreted as the difference of pressure along the
x and y and z axis. Note that the energy-momentum tensor~\eqref{eq:Tmunu-de}
is the most general one compatible with the metric~\eqref{metric}~\cite{2008ApJ...679....1K}.
Two quantities are introduced to define the degree of anisotropic
expansion: \begin{equation}
\begin{aligned}R & \,\equiv\,(\dot{a}/a-\dot{b}/b)/H\;=\;\Sigma_{x}-\Sigma_{y}\,,\\
S & \,\equiv\,(\dot{a}/a-\dot{c}/c)/H\;=\;2\Sigma_{x}+\Sigma_{y}\,.\end{aligned}
\label{dom}\end{equation}
 The reason why the cosmic parallax is allowed to be few orders of
magnitude larger than the one in \cite{2009arXiv0905.3727F} is based
on the presence of this anisotropic source arising after decoupling.
In particular, the value $\delta=-0.1$ is not completely excluded
by supernovae data, since it lies on the 2$\sigma$ contours of the
gamma-delta plane, if a prior on $w$ and $\Omega_{m}$ is assumed
\cite{2008ApJ...679....1K}. More phantom equation of state parameters
and/or larger matter densities allow for larger value of delta. In
addition, and more in general, time dependent delta and gamma functions,
mimicking for example specific minimally coupled vector field with
double power law potential, can escape these constraints. Our purpose
is to use this parameterization to model a very late-time evolution
of the shear, which is the reason why we linearised the dynamical
solutions around the critical points as denoted in the following paragraphs.

Considering the generalized Friedmann equation, the continuity equations
for matter and dark energy and no coupling between the two fluids,
the derived autonomous system reads%
\footnote{Notice that in \cite{2008ApJ...679....1K} there is a spurious factor
$x$ in the phase-space equations (8).%
} \cite{2008JCAP...06..018K,2008ApJ...679....1K}: \begin{equation}
\begin{aligned}U'= & U(U-1)[\gamma(3+R-2S)\;+\,\delta(3-2R+S)\,+\,3(w-w_{m})]\\
S'= & \frac{1}{6}(9-R^{2}+RS-S^{2})\big\{S[U(\delta+\gamma+w-w_{m})+w_{m}-1]-6\,\gamma\, U\big\}\\
R'= & \frac{1}{6}(9-R^{2}+RS-S^{2})\big\{R[U(\delta+\gamma+w-w_{m})+w_{m}-1]-6\,\delta\, U\big\},\end{aligned}
\label{sys}\end{equation}
 where $U\equiv\rho_{{\rm DE}}/(\rho_{{\rm DE}}+\rho_{m})$ and the
derivatives are taken with respect to $\log(A)/3$. In what follows
we will consider for simplicity that $w_{m}=0$, i.e. pressureless
matter . System~(\ref{sys}) exhibits many different fixed points,
defined as the solutions of the system $S'=R'=U'=0$. Beside the Einstein-de
Sitter case ($R_{*}=S_{*}=U_{*}=0$), the most physically interesting
for our purposes are the dark energy dominated solution \begin{equation}
R_{*}\,=\,\frac{6\delta}{\delta+\gamma+w-1},\;\;\; S_{*}\,=\,\frac{6\gamma}{\delta+\gamma+w-1},\;\;\; U_{*}=1,\label{eq:de-domination}\end{equation}
 and the scaling solution \begin{equation}
\begin{aligned} & R_{*}\,=\,\frac{3\delta(\delta+\gamma+w)}{2(\delta^{2}-\delta\gamma+\gamma^{2})},\quad S_{*}\,=\,\frac{3\gamma(\delta+\gamma+w)}{2(\delta^{2}-\delta\gamma+\gamma^{2})},\;\;\; U_{*}\,=\,\frac{w+\gamma+\delta}{w^{2}-3(\gamma-\delta)^{2}+2w(\gamma+\delta)},\end{aligned}
\label{scal}\end{equation}
 in which $\rho_{{\rm DE}}/\rho_{m}=const.$, i.e., the fractional
dark energy contribution to the total energy density is constant.
The latter is positive if the numerator and the denominator in the
expression for $U^{*}$ are either both positive or both negative;
moreover we should ensure the condition $U*<1$. If the numerator
is positive then $w>-(\gamma+\delta)$ is required; the denominator
is then positive only if $w>w_{1}=-(\gamma+\delta)+\sqrt{(\gamma+\delta)^{2}+3(\gamma-\delta)^{2}}$,
with $w_{1}$ being positive. Hence the most interesting case is when
they both are negative, which translates into these conditions: 1)
$w<-(\gamma+\delta)$; 2) $w>w_{2}=-(\gamma+\delta)-\sqrt{(\gamma+\delta)^{2}+3(\gamma-\delta)^{2}}$,
where now $w_{2}$ is negative. There are other conditions one should
impose to such solutions, in particular stability: we refer the reader
to Ref.~\cite{2008ApJ...679....1K,2008JCAP...06..018K} for the details.
All the specific cases discussed below satisfy these conditions. For
simplicity we also assume $w=-1$ and $\gamma=0$.

The cosmic parallax constrains the anisotropy at present, when the
dark energy density is of order $74\%$, hence not yet in the final
dark energy dominant attractor phase~(\ref{eq:de-domination}). Therefore
it must be either on its way to such a stage or, alternatively, on
the scaling critical solution~(\ref{scal}). We discuss separately
these two alternative scenarios. In the scaling case, in order not
to produce a too long accelerated epoch in the past, we ensure that
we just entering the accelerated regime.

\begin{figure*}[t!]
 \includegraphics[width=14cm]{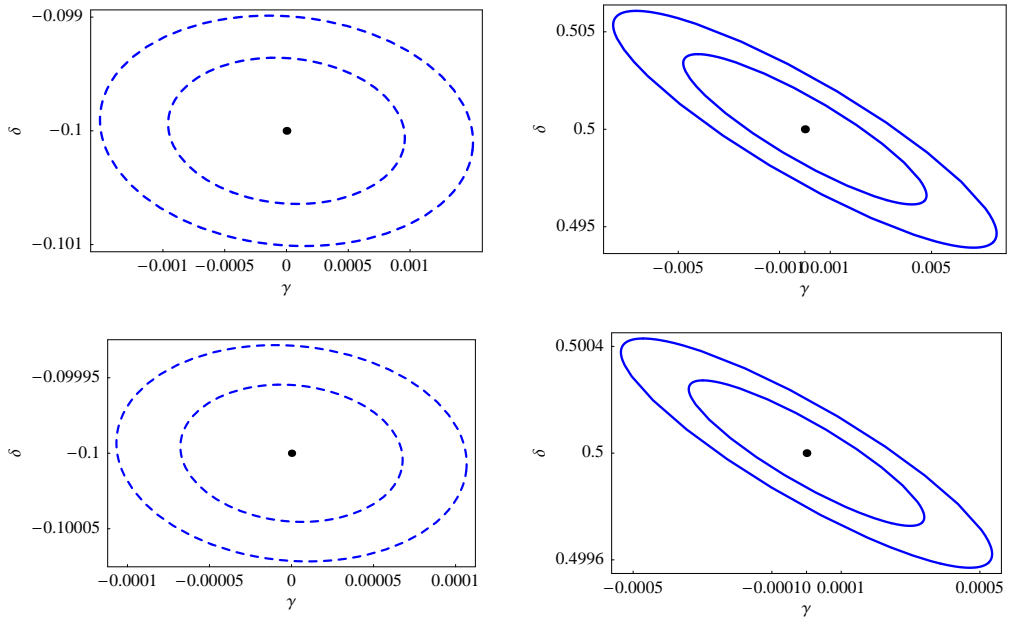}

\caption{Projected Fisher contours for the skewness dark energy parameters
for Gaia (upper panels) and Gaia+ specifications (lower panels). The
double contours identify $1\sigma$ and $2\sigma$ regions for $\Delta T=10$yrs.
The dashed lines represent the case of an ellipsoidal universe with
$\, w=-1$, $\, U_{0}=0.74\,$ and $\,\delta=-0.1\,$ ($R_{0}\simeq0.2$)
approaching the dark energy dominated critical point (where $\, U=1\,$
and $\, R_{*}\simeq0.3$), while the solid lines represent an ellipsoidal
universe that has just entered the scaling regime, with $\, w=-1$,
$\, U_{0}=0.74\,$ and $\,\delta=0.5\,$ ($R_{*}\simeq-0.5$). }

\label{DEcont} 
\end{figure*}



\begin{figure}[t]
 \includegraphics[width=9cm]{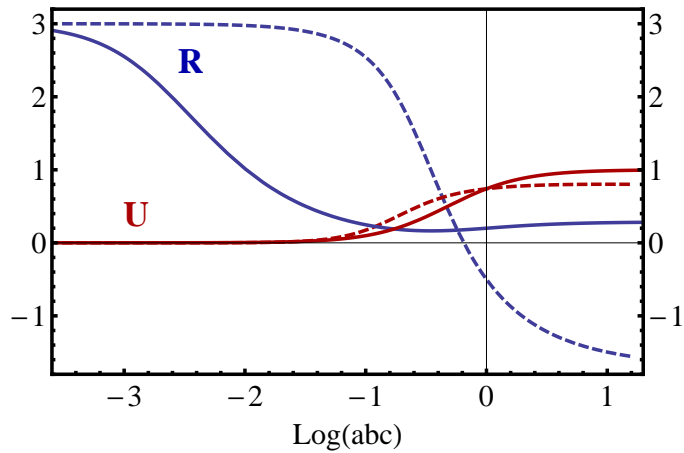}

\caption{Cosmological evolution of $R$ and $U$ for the two cases considered
(both with $w=-1$ and $U_{0}=0.74$): $\{\delta=-0.1\,,R_{0}\simeq0.2\}$
(solid lines) and $\{\delta=0.5\,,R_{0}\simeq-0.5\}$ (dashed lines).}

\label{back2} 
\end{figure}

We map our Fisher matrix~\eqref{fish} with the same experimental
specifications as in Table~\ref{gaia} into the new parameter space
${\bf p}=(\delta,\gamma)$. Our intention is to infer the order of
magnitude of the constraints this new cosmological tool would be able
to put on the dark energy skewness parameters. We apply the parameter
transformation $F'=A^{T}FA$, where $A_{ij}=\partial\Sigma_{0i}/\partial p_{j}$.
If we assume that the system has just entered the scaling solution,
the critical point~(\ref{scal}) approximately describes the dependence
of the dark energy anisotropy on the skewness parameters at present.
In this case we choose for the fiducial model $\delta=0.5$ and $\gamma=0$,
namely an ellipsoidal Universe with $R_{0}=-0.5$ (see Fig~\ref{back2})
and we choose initial conditions such that $U_{0}=0.74$. The final
attractor value is as expected rather close, $U_{0}=0.67$. The error
contours are shown in the right panels of Fig.~\ref{DEcont}, for
both Gaia and Gaia+ configurations.

Conversely, if the expansion is driven towards a future dark energy
dominated solution, equations~(\ref{eq:de-domination}) do not represent
the anisotropy parameters at present (see Fig~\ref{back2}). In order
to derive a more appropriate functional form for them, we solved the
linearized system~(\ref{sys}) around solution~(\ref{eq:de-domination})
and we fixed $\log(A)=0$ to select the present values. For this second
case, results are shown in the left panels of Fig.~\ref{DEcont}.
Notice that Fig~\ref{back2} depicts a late time expansion history
and aims just at illustrating the trend towards the critical points
from different values of the anisotropy parameters (namely $R=3$
in an earlier stage, but it might set to be vanishing at decoupling
by time dependent skewness parameters in specific models -- see \cite{2008ApJ...679....1K}).


%

The constraints on $\gamma$ and $\delta$ are in the range $10^{-3}\div10^{-4}$.
The current limits from SNIa data are then $2$ to $3$ orders of
magnitude weaker~\cite{2008ApJ...679....1K} and, even if the number
of supernovae will substantially increase in the near future, it might
be hard to improve the constraints at such a high level because of
the integral dependence of the luminosity distance on the skewness
parameters. Therefore the cosmic parallax seems to be an ideal candidate
for testing the anisotropically stressed dark energy.

The forecastings we presented in this section do not include possible
systematic effects. In our Fisher analysis we just took into account
the statistical errors. Several spurious effect must be considered
by the time real data is available. For instance, the peculiar velocity
of the objects need to be considered, although averaging it over a
large sample of uncorrelated objects it should be possible to eliminate
such a form of bias. Furthermore, this effect decreases with increasing
angular diameter distance to the object \cite{2008arXiv0809.3675Q}.
The main source of noise however could be due to the aberration change
induced by our own motion. Fortunately, both the aberration change
and the observer peculiar velocity signal have a dipolar signature
whereas cosmic parallax from Bianchi I models results in a superposition
of quadrupoles. Other minor effects like the temporal changes of local
lensing and microlensing (a parallax disturbances of few nanoarcseconds
are expected due to the weak microlensing \cite{2001MNRAS.323..952S}),
were not taken into account, as they require a more detail analysis,
beyond the scope of this paper.

\section{Discussions}

\label{disc} Any anisotropy will leave an imprint on the angular
distribution of objects that are able to trace cosmic expansion. If
such an anisotropy is present before the last scattering surface,
the CMB map will also be affected. The temperature field will carry
extra anisotropies mainly caused by the angular dependence of the
redshift at decoupling. By resolving geodesic equations and expanding
temperature anisotropies in spherical harmonics, it is straightforward
to relate the low multipole components to the eccentricities of the
model \cite{2008ApJ...679....1K}. In Bianchi I models the first notable
multipoles related to the CMB are the monopole and the quadrupole.
The observed value of the latter puts constraints on the shear \textit{at
last scattering} of order $10^{-5}$, taking into account the cosmic
variance. These constraints can be mapped into either magnetic field
\cite{2007PhRvD..76f3007C} or anisotropic dark energy limits, depending
on the source that gives rise to the anisotropy. In addition, one
expects the eccentricities to be non-vanishing if the expansion has
been somewhat anisotropic at decoupling. However it is in principle
also possible to escape detection from CMB if each scale factor has
expanded the same amount since last scattering, no matter how anisotropically.
Nonetheless, in all these analysis the anisotropy pattern is directly
added to the intrinsic standard FRW perturbations, a simplistic way
of treating the signal at large scale. In this still exploratory stage
of analysis it is worth stressing that CMB is indeed a very powerful
constraint on the shear at the time of decoupling, but with almost
no direct impact on late time expansion history. Complementary to
that, cosmic parallax, namely the temporal change of angular separation
of distant sources, is a direct and potentially powerful test of anisotropy
at small redshifts and \textit{at present}.

The anisotropic stress of dark energy is expected to have a leading
role in the generation of anisotropy at late times. It can be parameterized
by skewness parameters in the stress-energy tensor formulations, which
may be constant or time dependent functions. For example a minimally
coupled vector field satisfying quadrupole constraints was presented
in \cite{2008ApJ...679....1K}. The only way to test these models
is to use either the angular dependence of the magnitude or the angular
distribution of objects in the sky at recent time, i.e. either distant
source angular distribution or the real-time cosmic parallax, the
two relying on different techniques and having independent systematics
that complement each other.

We adopted a simple phenomenological model to describe late-time expansion
of the universe, filled by pressureless matter and an anisotropically
stressed dark energy component with two extra degrees of freedom,
namely the skewness parameters. No matter what the evolution of both
energy densities and shears was at high redshifts, we have shown that
Gaia will be able to constrain the skewness parameters up to $10^{-3}\div10^{-4}$
at 2$\sigma$, comparable to CMB tests at decoupling time, and 2$\div$3
orders of magnitude better than current supernovae Ia limits \cite{2008ApJ...679....1K}
(a Gaia+ experiment would improve them by one order of magnitude).

In this paper we discussed in detail the real-time technique; before
concluding we comment briefly on the possibility of testing anisotropy
through the accumulated effect on distant source (galaxies, quasars,
supernovae) distribution.

If sources shifts by as much as 0.1$\mu$as/year during the dark energy
dominated regime, then the accumulated shift will be of the order
of 1 arcmin in $10^{9}$years and up to fraction of a degree in the
time from the beginning of acceleration to now. If the initial distribution
is isotropic, this implies that sources in one direction will be denser
than in an orthogonal direction by roughly $1/90\approx10^{-2}$.
This anisotropy could be seen as a large-scale feature on the angular
correlation function of distant sources, where we expect any intrinsic
correlation to be negligible. The Poisson noise become negligible
for $N\gg10^{4}$: for instance, a million quasars could be sufficient
to detect the signal. Although the impact of the selection procedure
and galactic extinction is uncertain, this back-of-the-envelope calculation
shows that the real-time effect could be complemented by standard
large-scale angular correlation methods%
\footnote{We are indebted to an anonymous referee for this suggestion.%
}.

While finalizing this manuscript another work analysing the cosmic
parallax in Bianchi I models came out \cite{2009arXiv0905.3727F}.
Our work differs in many aspects. In \cite{2009arXiv0905.3727F} the
authors focused on the shear in models that isotropize, that is on
solutions of the shear dynamical equations that are decreasing function
of time. Hence the fact that they find a signal substantially lower
than ours is not surprising. In fact, in \cite{2009arXiv0905.3727F}
the background is described by a $\Lambda$CDM model where the non-FRW
quantities are driven by a constant equation of state. The analysis
is restricted to ellipsoidal universes, where the dependence on the
azimuthal angle is dropped and the signal is a pure quadrupole. Furthermore
in order to accomplish forecasting we have performed a Fisher Matrix
analysis of the signal contemplating two different experimental sets,
Gaia and Gaia+.

More in general, we have shown that, differently from LTB models with
off-centre observers \cite{2008arXiv0809.3675Q}, the cosmic parallax
signal in Bianchi I models is a combination of two quadrupole functions
of the two angular coordinates. Since the most important systematic
noises, caused by peculiar velocities and aberration changes, have
a dipolar functional form, Bianchi I models seem to be ideally testable,
though even in LTB models specific observational strategy aiming at
distinguishing the signal from the noise are possible \cite{2008arXiv0809.3675Q}.

Assuming that null geodesics are radial, we have provided an analytical
expression for the cosmic parallax in general Bianchi I models. This
assumption is motivated by a direct numerical calculation of the geodesic
for a source at redshift $z=1$ (see Appendix A).

CMB and cosmic parallax detect anisotropy at two different times and,
from an observational point of view, are completely independent on
each other: combining them together one will have the opportunity
to reconstruct the evolution of the anisotropy and test with high
accuracy the Copernican Principle.

\subsection*{Acknowledgements}

We would like to thank Tomi Koivisto for fruitful discussions.

\appendix

\section{Geodesic equations}

\label{geod} Restricting for simplicity to two dimensions, particularly
to the (X,Y) plane (where $\theta=\pi/2$), we now want to check whether
neglecting the curvature of null geodesic equations considerably affects
our results. Photons follow trajectories that are described by the
ensuing equations: \begin{eqnarray}
X'' & = & -2H_{X}t'p\label{geogeo1}\\
Y'' & = & -2H_{Y}t'q\label{geogeo2}\\
z' & = & -\frac{(1+z)}{t'}(a^{2}H_{X}p+b^{2}H_{Y}q),\label{geogeo3}\end{eqnarray}
 with the additional constraint $t'^{2}=a^{2}p+b^{2}q$ (note that
here $z$ is the redshift, not to be confused with the coordinate
Z). Here primes denote derivative with respect to the affine parameter
$\lambda$.

In order to solve system (\ref{geogeo1}-\ref{geogeo3}), the scale
factors as functions of time are required and hence one has to couple
to it the dynamical equations. Since we are integrating backward from
the observer position to the source location, typically at redshift
of order 1, we need to evaluate these functions in these redshift
range. We adopt the linearized solution of the dynamical system (\ref{sys})
in the vicinity of the critical point (\ref{eq:de-domination}): in
this way we take into account the effect of the shear arising from
an anisotropically stressed dark energy. The linearized equation for
the anisotropy parameters with $\delta=\gamma=-0.1$ and $w=-1$ are:
\begin{eqnarray}
S(A) & = & 0.27-0.40A^{-3.50}+0.38A^{-3.27}\label{lin}\\
R(A) & = & 0.27-0.07A^{-3.27}\\
U(A) & = & 1-0.26A^{-3.27}.\nonumber \end{eqnarray}

\begin{figure}[t]
 \includegraphics[width=9cm]{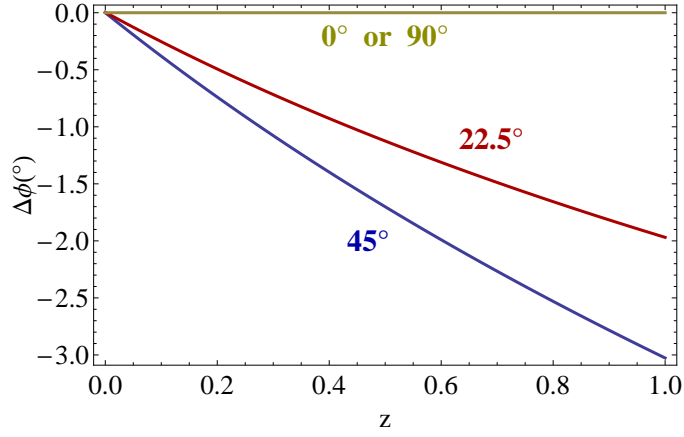}

\caption{Geodesic bending for a source at $z=1$ in the {}``$X$-$Y$'' plane
for the the dynamical system (\ref{sys}) in the vicinity of the critical
point (\ref{eq:de-domination}) in four different cases: photons arriving
at $\phi_{0}=90^{o}$ (along the $Y$-axis), $\phi_{0}=45^{o}$, $\phi_{0}=22.5^{o}$
and $\phi_{0}=0^{o}$ (along the $X$-axis). The total deviation is
always less than $7\%$, which validates the {}``straight geodesics''
approximation.}

\label{fig:geod-num} 
\end{figure}

The full set of equations with $S(A)$, $R(A)$ and $U(A)$ is then
:

\begin{eqnarray}
X' & = & p\qquad Y'=q\\
p' & = & -2H(A)\Big(\frac{S+3+R}{3}\Big)p\, t'\\
q' & = & -2H(A)\Big(\frac{S+3-2R}{3}\Big)q\, t'\\
z' & = & -\frac{(1+z)}{t'}H(A)\Big[a^{2}(A)\Big(\frac{S+3+R}{3}\Big)p+b^{2}(A)\Big(\frac{S+3-2R}{3}\Big)q\Big]\label{geo}\\
t'^{2} & = & a^{2}(A)p+b^{2}(A)q\\
a(A) & = & \exp{\Big[\int_{1}^{A}\Big(\frac{S+3+R}{3}\Big)\frac{dA'}{A}\Big]}\\
b(A) & = & \exp{\Big[\int_{1}^{A}\Big(\frac{S+3-2R}{3}\Big)\frac{dA'}{A}\Big]}\\
\Big(\frac{\dot{A}}{A}\Big)^{2}\Big[1-\frac{2}{9}(S^{2}+R^{2}-RS)\Big] & = & H_{0}^{2}(\Omega_{m0}A^{-3}+\Omega_{0DE}A^{-3(1+w+\delta+\gamma)}).\end{eqnarray}
 Results for a source located at z=1 are shown in Fig.~\ref{fig:geod-num}.

\bibliography{Bbian}
 \bibliographystyle{apsrev} 
\end{document}